\documentclass[12pt]{iopart}

\usepackage{amssymb}
\usepackage{iopams}
\usepackage{algorithm}
\usepackage{algorithmic}
\begin{document}

\title[Improvements on the security of arbitrated quantum signature protocols]{Improvements on the security of arbitrated quantum signature protocols}
\author{Zhiwei Sun$^1$, Ruigang Du$^1$, Banghai Wang$^2$, Qin Li$^3$ and Dongyang Long$^1$}
\address{$^1$School of Information Science and Technology, Sun Yat-sen University, 510006, P.R.China}
\address{$^2$Faculty of Computer, Guangdong University of
technology, Guangzhou 510006, P.R.China} 
\address{$^3$College of Information Engineering, Xiangtan University, Xiangtan 411105, P.R.China}
\ead{sunzhiwei1986@gmail.com,Duruigang@yahoo.com.cn,wangbanghai@gmail.com,
liqin805@163.com, issldy@mail.sysu.edu.cn}
%\date{\today}
\begin{abstract}
Recently, arbitrated quantum signature (AQS) used for signing quantum messages was proposed. It was claimed that the AQS schemes could guarantee unconditional security. However, in this paper, we show that all the presented AQS protocols are insecure. Due to the use of quantum one-time pad encryption, the signer Alice can always successfully acquire the receiver Bob's secret key and disavow any of her signatures. The detailed attack strategies and security analysis are described. Furthermore, the original versions of the protocols are revised and the security of the AQS protocols is improved accordingly. Besides, the presented method can also be against Alice's disavowal proposed by Gao et al. (Phys. Rev. A 84, 022344 (2011)).
\end{abstract}

\maketitle

\section{Introduction}
Digital signature schemes allow a signer Alice who has established a public key to sign a message in such a way that any other party who know the public key can verify that the message originated from signer and has not been modified in any way and the signer cannot repudiate it later \cite{2008Katz}. Digital signatures are commonly used for software distribution, financial transactions, and in other cases where it is important to detect forgery or tampering. However, digital signatures become increasingly vulnerable with more powerful quantum computation \cite{1994Shor,1996Grover} since their security is mostly based on the assumption of computational complexity. So, many scholars have begun to investigate quantum signature which is supposed to provide an alternative protocol with unconditional security. In 2002 Zeng and Keitel proposed an arbitrated quantum signature (AQS) which provides many merits \cite{2002Zeng}, and they announced that the unconditional security is ensured by using the correlation of Greenberger-Horne-Zeilinger (GHZ) triplet states and quantum one-time-pads \cite{Onetimepad2003}. In 2009 Li et al. \cite{2009Li} presented an AQS scheme using Bell states, which reduces the complexity of implementation by using Bell states instead of GHZ states. Recently, Zou et al further simplified this protocol achieving AQS without entangled state \cite{2010Zou}. Both of them still preserve the merits in Zeng et al's protocol.

Very recently Gao et al. show that these AQS protocols are not secure, and Bob can realize existential forgery of Alice's signature under known message attack \cite{2011Gao}. In this brief report, we will show that the AQS scheme is completely insecure if quantum one-time pad \cite{Onetimepad2003} is used, Alice can always obtain Bob's secret key and disavow all her signatures successfully. Having Bob's secret key, Alice has the ability to change her signature into any message in her favor after she has sent the signature to Bob. Therefore, some improvements are provided to enable the AQS schemes to circumvent our presented attack.

The remainder of this brief report is organized as follows. In Sec. \ref{sec2}, we analyze the security of the existing AQS protocols and present our attack. Then, in Sec. \ref{sec3}, we construct an AQS scheme similar to the scheme in Ref. \cite{2010Zou} which can stand against the presented attacks and the disavow attack in Ref. \cite{2011Gao}. The technique can also be used to improve the AQS scheme using entangled states \cite{2002Zeng,2009Li}. Finally, we give our conclusion.

\section{Security analysis for arbitrated quantum signature schemes} \label{sec2}

We first introduce quantum one-time-pad algorithm, which is helpful to understand our attack strategies. Then the AQS protocol using Bell states \cite{2009Li} and without entanglement \cite{2010Zou} are described briefly, and security analysis is demonstrated.

\subsection{A. Quantum one-time pad algorithm} \label{B}

For convenience, $E_{K}$ denotes the quantum one-time pad (QOTP) encryption \cite{Onetimepad2003} and the key is $K \in \{0, 1\}^{*}$, $K\geq 2n$. The QOTP encryption $E_{K}$ on the quantum message $|P\rangle = |p_{1}\rangle \otimes |p_{2}\rangle \otimes \cdots \otimes |p_{n}\rangle$ with $|p_{i}\rangle = \alpha_{i}|0\rangle + \beta_{i}|1\rangle$ can be described by
\begin{eqnarray}
|C\rangle = E_{K}|P\rangle =\bigotimes_{i=1}^{n} \sigma_{x}^{K^{2i}} \sigma_{z}^{K^{2i-1}} |p_{i}\rangle,
\end{eqnarray}
where $K^{j}$ denotes the $j$th bit of $K$, and $\sigma_{x}$ and $\sigma_{z}$ are Pauli operations. The corresponding decryption $D_{K}$ is
\begin{eqnarray}
D_{K}|C\rangle = \bigotimes_{i=1}^{n} \sigma_{z}^{K^{2i-1}} \sigma_{x}^{K^{2i}} |c_{i}\rangle,
\end{eqnarray}
where $|c_{i}\rangle$ denotes the $i$th qubit of the ciphertext $|C\rangle$.

\subsection{B. AQS scheme using Bell states} \label{Bell}
The AQS protocol using Bell states \cite{2009Li} is as follows.

\textit{Initializing phase.}

Alice and Bob share a key with the arbitrator through quantum key distribution protocols, i.e., $K_{A}$ and $K_{B}$ respectively, and they also share $n$ Bell states $|\phi^{+}\rangle = \frac{1}{\sqrt{2}}(|00\rangle + |11\rangle)_{AB}$, where the subscripts $A$ and $B$ correspond to Alice and Bob, respectively.

\textit{Signing phase.}

$S1.$ Alice obtains three copies of the quantum message $|P\rangle = \otimes_{i=1}^{n}|p_{i}\rangle$ to be signed.

$S2.$ Using the key $K_{A}$, Alice transforms one copy of $|P\rangle$ into $|R_{A}\rangle$, i.e., $|R_{A}\rangle = M_{K_{A}}|P\rangle$. We notice that $M_{K_{A}}$ denotes a unitary operator, and it may be either commutative or non-commutative with other quantum operators. In Ref. \cite{2008Zeng}, the author gave an example to show how the quantum state $|R_{A}\rangle$ is generated by Alice,
\begin{eqnarray} \label{eqma}
|R_{A}\rangle &=& M_{K_{A}}|P\rangle \nonumber \\
&=& \bigotimes_{i=1}^{n} M_{K_{A}^{i}}|p_{i}\rangle = \bigotimes_{i=1}^{n} \sigma_{x}^{1\oplus{K_{A}^{i}}} \sigma_{z}^{K_{A}^{i}} |p_{i}\rangle,
\end{eqnarray}
where $K_{A}^{i}$ is the $i$th bit of $K_{A}$, but it does not mean that Eq. (\ref{eqma}) is the only format of $M_{K_{A}}$. The purpose of this example is to present a detailed mathematical formulation of generating the state $R_{A}$. As Gao et al. has shown that if $M_{K_{A}}$ is commutative with other quantum operators, existential forgery attack is demonstrated \cite{2011Gao}. So non-commutative property should be included in $M_{K_{A}}$ \cite{2011Jeong}.

$S3.$ Alice combines each qubit in the second copy of $|P\rangle$ and the Bell state by carrying out a joint measurement on both states and obtains the three-particle entangled state,
\begin{eqnarray}
|\phi_{i}\rangle &=& |p_{i}\rangle\otimes|\phi^{+}_{i}\rangle \nonumber \\
   &=& \frac{1}{2}\{|\phi^{+}\rangle_{A}(\alpha_{i}|0\rangle+\beta_{i}|1\rangle)_{B}+|\phi^{-}\rangle_{A}(\alpha_{i}|0\rangle-\beta_{i}|1\rangle)_{B}\nonumber \\
   &&+ |\psi^{+}\rangle_{A}(\alpha_{i}|1\rangle+\beta_{i}|0\rangle)_{B} + |\psi^{-}\rangle_{A}(\alpha_{i}|1\rangle-\beta_{i}|0\rangle)_{B} \},\nonumber \\
\end{eqnarray}
where $|\phi^{+}\rangle_{A}$, $|\phi^{-}\rangle_{A}$, $|\psi^{+}\rangle_{A}$ and $|\psi^{-}\rangle_{A}$ are the four Bell states \cite{Bellstates}.
Then she implements a Bell measurement on each three-particle entangled state $|\phi_{i}\rangle$, obtaining the measurement result $|\mathcal{M}_{A}\rangle = \otimes_{i=1}^{n} |\mathcal{M}_{A}^{i}\rangle$, where $|\mathcal{M}_{A}^{i}\rangle$ are random Bell states. The role of $|\mathcal{M}_{A}\rangle$ is to help Bob to retrieve the second copy of message $|P\rangle$ by teleportation via Bell states previously shared between them.

$S4.$ Alice generates the signature $|S\rangle = E_{K_{A}}(|\mathcal{M}_{A}\rangle, |R_{A}\rangle)$ of message $|P\rangle$ by encrypting $|\mathcal{M}_{A}\rangle$ and $|R_{A}\rangle$ with the secret key $K_{A}$.

$S5.$ Alice transmits the signature $|S\rangle$ and the third copy of message $|P\rangle$ to Bob.

\textit{Verifying phase.}

$V1.$ Bob encrypts $|S\rangle$ and $|P\rangle$ using the key $K_{B}$, obtaining $|Y_{B}\rangle = E_{K_{B}}( |S\rangle, |P\rangle)$, and sends it to the arbitrator.

$V2.$ The arbitrator decrypts the received ciphertext $|Y_{B}\rangle$ with $K_{B}$ and $K_{A}$, getting $|\mathcal{M}_{A}\rangle$, $|R_{A}\rangle$ and $|P\rangle$. Then the arbitrator sets the verification parameter $V=1$ if $|R_{A}\rangle = M_{K_{A}}|P\rangle$; otherwise he sets $V=0$. Quantum state comparison was discussed in detail in Ref. \cite{2009Li}.

$V3.$ The arbitrator can recover $|S\rangle$ and $|P\rangle$ as the compared states can be recovered after the comparison if they are indeed equal. As $|\mathcal{M}_{A}\rangle$ are Bell states, it can be distinguished and replicated many copies. Then he sends the encrypted results $|Y_{TB}\rangle = E_{K_{B}}(|\mathcal{M}_{A}\rangle, |S\rangle, |P\rangle, V)$ to Bob.

$V4.$ Bob decrypts the received $|Y_{TB}\rangle$ and judges whether $V=1$. If not, he considers that the signature is forged and stops the protocol.

$V5.$ According to $|\mathcal{M}_{A}\rangle$, Bob can restore the second copy of $|P\rangle$ via teleportation by Alice. Then he compares it with the copy received from the arbitrator and accepts the signature when they are equal; otherwise he considers that the signature has been forged and rejects it.

\subsection{C. AQS scheme without entanglement}
In 2010, Zou et al. pointed out that the AQS scheme using Bell states can be repudiated by the receiver Bob, and they improved the AQS scheme by using a public board to conquer this shortcoming. Only the following two things are needed to do:

$(1)$. In the signing phase, Alice first chooses a random number $r \in \{0, 1\}^{2n}$ and transforms all $|P\rangle$ into secret qubit strings $|P^{'}\rangle = E_{r}(|P\rangle)$. Then they use $|P^{'}\rangle$ instead of $|P\rangle$ in all following steps.

$(2)$. In the verifying phase, Bob informs Alice by the public board to publish $r$ after he finished his verifying. Then, Alice publishes $r$ by the public board. Finally, Bob gets back $|P\rangle$ from $|P^{'}\rangle$ by $r$ and holds $(|S\rangle, r)$ as Alice's signature for the quantum message $|P\rangle$.

In Ref. \cite{2010Zou}, the author also said that in order to achieve a higher efficiency in transmission, they do the following improvement:

$(3)$. In step $V1$, Bob does not send his measuring result $|\mathcal{M}_{A}\rangle$ to the arbitrator, and the arbitrator need not send it back. In addition, the arbitrator informs Alice and Bob by the public board to abort the scheme if he found the signature being forged.

Zou's AQS scheme without using entanglement \cite{2010Zou} is as follows.

\textit{Initializing phase}

Three keys $K_{AB}$, $K_{A}$ and $K_{B}$ are shared between Alice and Bob, Alice and the arbitrator, Bob and the arbitrator respectively.

\textit{Signing phase}

$S1$. Alice obtains three copies of the quantum message $|P\rangle = \otimes_{i=1}^{n}|p_{i}\rangle$, and encrypts each of them into $|P^{'}\rangle$ using a random number $r$ as the key.

$S2$. Alice performs the following encryptions $|R_{AB}\rangle = E_{K_{AB}}|P^{'}\rangle$, $|S_{A}\rangle = E_{K_{A}}|P^{'}\rangle$, and sends $|P^{'}\rangle$, $|R_{AB}\rangle$ and $|S_{A}\rangle$ to Bob.

\textit{Verifying phase}

$V1$. Bob sends $|Y_{B}\rangle = E_{K_{B}}(|P^{'}\rangle, |S_{A}\rangle)$ to the arbitrator.

$V2$. The arbitrator decrypts $|Y_{B}\rangle$ and verifies whether $|S_{A}\rangle = E_{K_{A}}|P^{'}\rangle$. If the equation holds, he sets the verification parameter $V_{T} =1$; otherwise he sets $V_{T}=0$. He announces the verification parameter $V_{T}$ by the public board and regenerates $|Y_{B}\rangle$ and sends it back to Bob.

$V3$. If $V_{T}=0$, Bob rejects the signature; otherwise he decrypts $|Y_{B}\rangle$ and verifies whether $E_{K_{AB}}|P^{'}\rangle = |R_{AB}\rangle$. If $E_{K_{AB}}|P^{'}\rangle = |R_{AB}\rangle$, he sets the verification parameter $V_{B} = 1$; otherwise he sets $V_{B} = 0$. He announces the verification parameter $V_{B}$ by the public board.

$V4$. If $V_{B} = 1$, Alice publishes $r$ by the public board, and Bob gets back $|P\rangle$ from $|P^{'}\rangle$ by $r$ and stores $(|S_{A}\rangle, r)$ as Alice's signature for the quantum message $|P\rangle$.

\subsection{D. Security analysis of the AQS schemes}
In the remainder of this subsection, we'll show that the AQS schemes are insecure. Because Alice can obtain Bob's secret key and deny her signature successfully by the property of QOTP encryption \cite{Onetimepad2003}.

\subsubsection{1. Alice's general attack on AQS scheme without entanglement}
We describe Alice's general attack in detail in the following.

\begin{table}
\caption{\label{tab:special} Relations of Alice's key $K_{A}$, $|\mathcal{M}_{A}\rangle$ and $E_{K_{A}}\otimes I|\mathcal{M}_{A}\rangle$}
\begin{indented}
\item[]\begin{tabular}{@{}lllll}
\br
$K_{A}\backslash|\mathcal{M}_{A}\rangle$ & $|\phi^{+}\rangle$ & $|\phi^{-}\rangle$ & $|\psi^{+}\rangle$ & $|\psi^{-}\rangle$\\
\mr
$00$ & $|\phi^{+}\rangle$ & $|\phi^{-}\rangle$ & $|\psi^{+}\rangle$ & $|\psi^{-}\rangle$\\
$01$ & $|\phi^{-}\rangle$ & $|\phi^{+}\rangle$ & $|\psi^{-}\rangle$ & $|\psi^{+}\rangle$\\
$10$ &  $|\psi^{+}\rangle$ & $-|\psi^{-}\rangle$ & $|\phi^{+}\rangle$ & $-|\phi^{-}\rangle$\\
$11$ & $-|\psi^{-}\rangle$ & $|\psi^{+}\rangle$ & $-|\phi^{-}\rangle$ & $|\phi^{+}\rangle$\\
\br
\end{tabular}
\end{indented}
\end{table}

Her attack begins in step $S2$. Alice prepares an ordered $n$ Bell states $|\phi^{+}\rangle = \frac{1}{\sqrt{2}}(|00\rangle + |11\rangle)_{TH}$, where the subscripts $T$ and $H$ denote different particles. We denote the $n$ ordered Bell states with $(T_{1}, H_{1}), (T_{2}, H_{2}), (T_{3}, H_{3}), \cdots, (T_{n}, H_{n})$, where the subscripts indicates the pair order in the sequence. Alice takes one particle from each Bell state to form an ordered particle sequence which is denoted by $|T\rangle = (T_{1}, T_{2}, T_{3}, \cdots, T_{n})$. The remaining particles compose another particle sequence $|H\rangle = (H_{1}, H_{2}, H_{3}, \cdots, H_{n})$. Alice performs the following encryptions $|R_{AB}\rangle = E_{K_{AB}}|P^{'}\rangle$, $|S_{A}\rangle = E_{K_{A}}|P^{'}\rangle$, and sends $|T\rangle$, $|R_{AB}\rangle$ and $|S_{A}\rangle$ instead of $|P^{'}\rangle$, $|R_{AB}\rangle$ and $|S_{A}\rangle$ to Bob. As Alice transforms the quantum message $|P\rangle$ into $|P^{'}\rangle$ using a random number $r$, $|P^{'}\rangle$ will be known to nobody. Furthermore, non-orthogonal states can't be reliably distinguished. Therefore, Bob won't notice Alice's attack and accepts $|T\rangle$ as the signed message.

In the verifying phase, Bob sends $|Y_{B}\rangle = E_{K_{B}}(|T\rangle, |S_{A}\rangle)$ to the arbitrator for verification.
Alice intercepts it, obtaining $E_{K_{B}}|T\rangle$. Then Alice can learn Bob's secret key $K_{B}$ exactly by performing Bell-basis measurement on $E_{K_{B}}|T\rangle$ and $|H\rangle$ simultaneously, which can refer to TABLE \ref{tab:special} and Ref. \cite{Onetimepad2003}. For example, if $(T, H) = |\phi^{+}\rangle$ and $(E_{K_{B}}T, H) = |\psi^{+}\rangle$, the secret key $K_{B} = 10$.

Alice generates $|Y_{B}^{'}\rangle = E_{K_{B}}(|P^{'}\rangle, |S_{A}\rangle)$ by encrypting $|P^{'}\rangle$ and $|S_{A}\rangle)$ using the key $K_{B}$, and sends it to the arbitrator.
When the arbitrator announces the verification parameter $V_{T}=1$ by the public board and sends $|Y_{B}^{'}\rangle = E_{K_{B}}(|P^{'}\rangle, |S_{A}\rangle)$ back to Bob in step $V2$, Alice intercepts it.
Then Alice randomly selects a quantum message $|P^{"}\rangle$ where $|P^{"}\rangle \neq |P^{'}\rangle$ and generates $|S^{'}\rangle = E_{K_{A}}|P^{"}\rangle$. Then Alice sends $|Y_{B}^{"}\rangle = E_{K_{B}}(|P^{'}\rangle, |S_{A}^{'}\rangle)$ to Bob.
Bob will accept this signature without noticing Alice's attack in step $V3$ and $V4$. When dispute appears Bob requires to make a judgment by providing $(|P\rangle, |S^{'}\rangle, r)$ to the arbitrator. Then the arbitrator generates $|P^{'}\rangle$ by encrypting $|P\rangle$ using $r$, and verifies whether $|S^{'}\rangle = E_{K_{A}}|P^{'}\rangle$. Obviously the modified signature will not pass verification, and hence Alice denies having signed the message successfully.

\subsubsection{2. Alice's general attack on AQS schemes with entanglement}
Now, we show that the above attack is more powerful than the presented attack in Ref. \cite{2011Gao}. Because the AQS schemes using entanglement \cite{2002Zeng,2009Li} are total insecure with the above attack strategy, i.e., Alice can completely obtain Bob's secret key and change her signature for any message in her favor, which is described explicitly as follows.

We take the AQS scheme using Bell states as an example. Similar to the above attack, Alice prepares an ordered $4n$ Bell states $|\phi^{+}\rangle = \frac{1}{\sqrt{2}}(|00\rangle + |11\rangle)_{TH}$, where the subscripts $T$ and $H$ denote different particles. We denote the $4n$ ordered Bell states with $(T_{1}, H_{1}), (T_{2}, H_{2}), (T_{3}, H_{3}), \cdots, (T_{4n}, H_{4n})$, where the subscripts indicates the pair order in the sequence. Alice takes one particle from each Bell state to form an ordered particle sequence which is denoted by $|T\rangle = (T_{1}, T_{2}, T_{3}, \cdots, T_{4n})$. The remaining particles compose another particle sequence $|H\rangle = (H_{1}, H_{2}, H_{3}, \cdots, H_{4n})$. Then she send the $|T\rangle = (T_{1}, T_{2}, T_{3}, \cdots, T_{4n})$ as the signature and the signed message instead of $|S\rangle$ and $|P\rangle$ to Bob in step $S5$. In the verifying Phase, Bob encrypts $|T\rangle = (T_{1}, T_{2}, T_{3}, \cdots, T_{4n})$ using the $K_{B}$ obtaining $|Y_{B}\rangle$ and sends it to the arbitrator. Then Alice can completely access Bob's secret key $K_{B}$ using the similar method as described above. After having Bob's secret key, Alice has the ability to change her signature into any message. Alice chooses the message $|P^{'}\rangle$ in her favor and generates the signature $|S^{'}\rangle$, and encrypts them using Bob's secret key, i.e., $|Y_{B}^{'}\rangle = E_{K_{B}}(|P^{'}\rangle, |S^{'}\rangle)$. Then she sends $|Y_{B}^{'}\rangle$ to the arbitrator. It is easy to see that Bob will accept $|P^{'}\rangle$, $|S^{'}\rangle$ as a valid signature.

It should be stressed that the security of the arbitrator quantum signature is based on the chosen symmetric-key encryption scheme and the shared secret key of the participants. The reasons of our presented attack more powerful than the Gao's attack \cite{2011Gao} are: on the one hand, if Alice obtains Bob's secret key, then Alice can change her mind about the message $|P\rangle$ in her favor, which is not fair to Bob and not allowed in classical digital signature;
on the other hand, if the secret key is obtained by Eve, the scheme will be totally insecure in cryptography, which is also known as total break, i.e., Alice can forge signatures for any message, while Gao's attack is only existential forgery attack \cite{2011Gao}. Furthermore, in step $S2$ of AQS scheme using Bell states, we have pointed out that the existential forgery attack will not exist if non-commutative property is included in $M_{K_{A}}$ \cite{2011Jeong}.

\section{A secure AQS scheme without using entangled states} \label{sec3}
We have analyzed that the existing AQS schemes \cite{2002Zeng,2009Li,2010Zou} are totally insecure. And from Ref. \cite{2010Zou}, we know that the AQS scheme without using entangled states reduces the complexity of implementing the scheme and maintains all other merits of AQS scheme using Bell states \cite{2009Li} and the AQS scheme using GHZ states \cite{2002Zeng}. Therefore, in this section, we present a new AQS scheme without using entangled states that can avoid above attack and preserves all the merits of AQS scheme of Ref. \cite{2010Zou}.

As known to all that a secure arbitrated quantum signature should satisfy two conditions: one is that the signature should not be forged by the attacker (including the malicious receiver and the not fully trusted arbitrator) and the other is the impossibility of disavowal by the signatory and the receiver. In the AQS scheme without using entangled states \cite{2010Zou}, Alice sends $|P^{'}\rangle$, $|R_{AB}\rangle$ and $|S_{A}\rangle$ to Bob in step $S2$. $|R_{AB}\rangle$ is used to prevent the arbitrator from forge Alice's signature as he is not fully trusted by Alice and Bob, and $K_{AB}$ is kept secret from him; $|S_{A}\rangle$ is used for avoiding Bob forging her signature as he does not have Alice's secret key $K_{A}$, meanwhile, the secret key $K_{A}$ is included in $|S_{A}\rangle$ which will also prevent Alice from repudiating her signature; $|P^{'}\rangle$ is used to avoid being disavowed by Bob.Therefore, $|P^{'}\rangle$, $|R_{AB}\rangle$ and $|S_{A}\rangle$ are essential to achieve a secure signature.

Alice is able to deny her signature and obtain Bob's secret key because Bob has not verified the validity of the signed message $|P^{'}\rangle$ in step $V1$. We notice that $|P^{'}\rangle$ may be replaced by some entangled states, and Alice obtains Bob's secret key by the property of quantum one-time pad, so she can deny her signature and change the original message into any other message in her favor. To avoid being disavowed and forged by Alice, Bob must verify the validity of the signed message $|P^{'}\rangle$ before he sends it to the arbitrator in step $V1$.

Gao et al. \cite{2011Gao} showed that Alice can disturb the signature $|S_{A}\rangle$ when the arbitrator sends the $|Y_{B}\rangle$ back to Bob in step $V2$. As only $|S_{A}\rangle$ is modified by Alice and $|S_{A}\rangle$ is not useful for Bob's verification in step $V3$, Bob will accept this signature as a valid one. However, when dispute appears, Alice can always successfully disavow her signature because the disturbed signature will not pass the arbitrator's verification.
However, Gao's attack is actually a special DOS attack which is inevitable in all existing protocols. And one of important property of the quantum signature is to ensure the integrity (or authenticity) of transmitted quantum messages. Adding a quantum message authentication to the quantum signature protocol \cite{2011Gao} may be not a proper way for secure quantum signature.

We notice that $|S_{A}\rangle$ is completely indistinguishable to Bob, when he receives the modified $|S_{A}\rangle$ he cannot verify its validity. That is the main reason why Alice's disavow is always successful. To avoid Gao's attack, Bob need authenticate the validity of the $|S_{A}\rangle$ when he receives it back from the arbitrator. We give a simple method to verify the validity of the the $|S_{A}\rangle$ when Bob receives it back from the arbitrator, thereby avoiding Gao's attacks.
The AQS scheme is specified in the following.

\textit{Initializing phase}

Three keys $K_{AB}$, $K_{A}$ and $K_{B}$ are shared between Alice and Bob, Alice and the arbitrator, Bob and the arbitrator respectively. The lengths of these keys depend on the chosen cryptographic algorithms in the signing and verifying phases.

\textit{Signing phase}

$S1$. Alice obtains four copies of the quantum message $|P\rangle = \otimes_{i=1}^{n}|p_{i}\rangle$, and transforms each of them into $|P^{'}\rangle$ using a random number $r$ as the key, i.e., $|P^{'}\rangle = M_{r}|P\rangle$ where $M_{r}$ is a non-commutative unitary operator.

$S2$. Alice performs the following encryptions $|R_{AB}\rangle = E_{K_{AB}}|P^{'}\rangle$, $|S_{A}\rangle = E_{K_{A}}|P^{'}\rangle$, $|S_{A}\rangle = E_{K_{A}}|P^{'}\rangle$ and sends $|P^{'}\rangle$, $|R_{AB}\rangle$, $|S_{A}\rangle$ and $|S_{A}\rangle$ to Bob. Note that there are two copies of $|S_{A}\rangle$ (in order to facilitate the expression, we denote them $|S_{A}\rangle_{1}$ and $|S_{A}\rangle_{2}$ respectively), one $|S_{A}\rangle_{1}$ for the arbitrator to verify the validity of the signature, the other $|S_{A}\rangle_{2}$ is used to against Alice's disavowal proposed by Gao et al. \cite{2011Gao}.

\textit{Verifying phase}

$V1$. Before Bob sending $|Y_{B}\rangle = E_{K_{B}}(|P^{'}\rangle, |S_{A}\rangle_{1})$ to the arbitrator for verification, he verifies the validity of the  $|P^{'}\rangle$ and $|S_{A}\rangle$ first. If $|R_{AB}\rangle = E_{K_{AB}}|P^{'}\rangle$ and the two copies of $|S_{A}\rangle$ are identical, he believes that Alice is honest and send $|Y_{B}\rangle$ to the arbitrator; otherwise, he terminates the protocol and rejects the signature. Note that Bob keeps one copy of $|S_{A}\rangle$ in his hand and it will be used to verify whether the signature sent to the arbitrator is modified by Eve.

$V2$. The arbitrator decrypts $|Y_{B}\rangle$ and verifies whether $|S_{A}\rangle_{1} = E_{K_{A}}|P^{'}\rangle$. If the equation holds, he sets the verification parameter $V_{T} =1$; otherwise he sets $V_{T}=0$. He announces the verification parameter $V_{T}$ by the public board and regenerates $|Y_{B}\rangle$ and sends it back to Bob.

$V3$. If $V_{T}=0$, Bob rejects the signature; otherwise he decrypts $|Y_{B}\rangle$ and verifies whether $E_{K_{AB}}|P^{'}\rangle = |R_{AB}\rangle$ and $|S_{A}\rangle_{1}$ = $|S_{A}\rangle_{2}$. If both of them hold, he sets the verification parameter $V_{B} = 1$; otherwise he sets $V_{B} = 0$. He announces the verification parameter $V_{B}$ by the public board.

$V4$. If $V_{B} = 1$, Alice publishes $r$ by the public board, then Bob gets back $|P\rangle$ from $|P^{'}\rangle$ by $r$ and he accepts the signature $|S_{A}\rangle$ of the message $|P\rangle$ and stores $(|P\rangle, |S_{A}\rangle, |S_{A}\rangle)$ for resolving disputes when Alice disavows her signature.

According previous analysis, the new specific AQS scheme without entangled states is secure, i.e., it is secure against our presented attack and Gao's attack \cite{2011Gao} and maintains all the merits of the existing AQS schemes \cite{2002Zeng,2009Li,2010zou}. Furthermore, we give a specific protocol while Gao just give a possible improvement method and our protocol is easier to be implemented than the improved AQS scheme with quantum message authentication of Ref. \cite{2011Gao}. Similarly, the AQS schemes using entangled states \cite{2002Zeng,2009Li} can also be improved by the above method.

\section{Conclusion}
In this brief report, we present a general attack which shows that the existing AQS protocols \cite{2002Zeng,2009Li,2010Zou} are insecure because Alice can obtain Bob's secret key and disavow any message she ever signed and forge signatures for any message in her favor. And we improve the AQS schemes to against all the attacks. To avoid Alice obtaining Bob's secret key, Bob must firstly verify the validity of the signed message before he sends it to the arbitrator. Note that our presented AQS scheme, on the one hand, can avoid being disavowed and forged by malicious Alice, on the other hand, preserves all merits of the existing schemes \cite{2002Zeng,2009Li,2010Zou}.

 \ack \indent This work is in part supported by the Key Project of NSFC-Guangdong
Funds (No.U0935002).


\section*{References}
\begin{thebibliography}{10}

\bibitem{2008Katz} Katz, Jonathan and Lindell, Yehuda 2007 Introduction to Modern Cryptography \emph{Chapman \& Hall/Crc Cryptography and Network Security Series}

\bibitem {1994Shor} P. W. Shor 1994 Algorithm for quantum computation: Discrete logarithm and factoring algorithm \emph{Proceedings of the 35th Annual
Symposium on Foundations of Computer Science} (IEEE Computer Soceity
Press, Los Alamos, CA) 124

\bibitem{1996Grover} Lov K. Grover1996  A fast quantum mechanical algorithm for database search \emph{In Proceedings of 28th Annual ACM Symposium on Theory of Computing} (New York) 212-219
\bibitem{2002Zeng} Zeng, Guihua  and Keitel, Christoph H. 2002 Arbitrated quantum-signature scheme \emph{Phys. Rev. A} {\bf 65} 042312
\bibitem{Onetimepad2003} Boykin, P. Oscar and Roychowdhury, Vwani 2003 Optimal encryption of quantum bits \emph{Phys. Rev. A} {\bf 67} 042317
\bibitem{2009Li} Li, Qin  and Chan, W. H. and Long, Dong-Yang 2009 Arbitrated quantum signature scheme using Bell states \emph{Phys. Rev. A} {\bf 79} 054307.
\bibitem{2010Zou} Zou, Xiangfu  and Qiu, Daowen 2010 Security analysis and improvements of arbitrated quantum signature schemes \emph{Phys. Rev. A} {\bf 82} 042325
\bibitem{2011Gao} Gao, Fei and Qin, Su-Juan and Guo, Fen-Zhuo and Wen, Qiao-Yan 2011 Cryptanalysis of the arbitrated quantum signature protocols \emph{Phys. Rev. A} {\bf 84} 022344
\bibitem{2008Zeng} Zeng, Guihua 2008 Reply to ``Comment on `Arbitrated quantum-signature scheme' '' \emph{Phys. Rev. A} {\bf 78} 016301
\bibitem{2011Jeong} Jeong Woon Choi, Ku-Young Chang, Dowon Hong 2011 \emph{arXiv:1106.5318v1}
\bibitem{Bellstates} Kwiat, Paul G. and Mattle, Klaus  and Weinfurter, Harald  and Zeilinger, Anton  and Sergienko, Alexander V. and Shih, Yanhua 1995 New High-Intensity Source of Polarization-Entangled Photon Pairs \emph{Phys. Rev. Lett.} {\bf 75} 4337--4341
 
\end{thebibliography}
\end{document}